\documentclass[1p,times,11pt]{elsarticle}

\usepackage{amssymb}

\usepackage{amsfonts}
\usepackage{amssymb}
\usepackage{amsthm,amsmath}
\usepackage{mathrsfs}
\usepackage{indentfirst}
\usepackage{multirow}
\usepackage[linesnumbered,ruled,vlined]{algorithm2e}
\usepackage[section]{placeins}
\biboptions{sort&compress}

\usepackage{lineno}

\journal{Nano Commun.Netw.}

\begin{document}
	
	\begin{frontmatter}
		
		
		
		\title{An Energy Balance Cluster Network Framework Based on Simultaneous Wireless Information and Power Transfer}
		\tnotetext[t1]{This work was supported by the National Natural Science Foundation of China under Grant 61202384 and 61971314.}
		
		
		\author[address1]{Juan Xu\corref{cor1} }
		\ead{jxujuan@tongji.edu.cn}
		\cortext[cor1]{corresponding author}
		
		\author[address1]{Ruofan Wang}
		\ead{wrfbwcx@163.com}
		
		\author[address1]{Yan Zhang}
		\ead{1830732@tongji.edu.cn}
		
		\author[address1]{Hongmin Huang}
		\ead{freeastime@163.com}
		
		\address[address1]{College of Electronic and Information Engineering, Tongji University, 201804, Shanghai, China}

		\begin{abstract}
			
			Wireless NanoSensor Network (WNSN) is a brand-new type of sensor network with broad application prospects. In view of the limited energy of nano-nodes and unstable links in WNSNs, we propose an energy balance cluster network framework (EBCNF) based on Simultaneous Wireless Information and Power Transfer (SWIPT). The EBCNF framework extends the network lifetime of nanonodes and uses a clustering algorithm called EBACC (an energy balance algorithm for intra-cluster and inter-cluster nodes) to make the energy consumption of nodes more uniform. Simulation shows that the EBCNF framework can make the network energy consumption more uniform, reduce the error rate of data transmission and the average network throughput, and can be used as an effective routing framework for WNSNs.
		\end{abstract}


	\end{frontmatter}

	\section{Introduction}
	Due to the development of  nanotechnology and the emergence of new materials like controlling materials from one nanometer to several hundred nanometers , the realization and application of Wireless Nano Sensor Networks (WNSNs) is feasible\cite{1}. Due to the extremely limited storage capacity of nano batteries, the communication performance of WNSNs is limited \cite{2,3}. Therefore, energy harvesting has always been a research focus of WNSNs. Obtaining energy from the surrounding environment provides a promising method for improving the network lifetime and performance of energy-constrained WNSN \cite{3}. Since electromagnetic signals not only carry information but also energy, a method for processing environmental electromagnetic signal information while collecting energy is proposed \cite{4}. For WNSN, SWIPT(Simultaneous Wireless Information and Power Transfer) is different from piezoelectric energy harvesting systems. It can provide stable energy for nano-nodes and is a promising charging method. In traditional electromagnetic communication, Varshney first proposed the idea of transmitting power and information at the same time \cite{4}. Grover et al. analyzed SWIPT based on frequency selective channel which provides ideas for simultaneous transmission of power and information on the THz band \cite{5}. Taking into account the non-linearity of the rectifier, Bruno proposed a non-linear rectifier model for SWIPT technology \cite{6}. Considering that SWIPT can overcome the energy bottleneck of nano sensor networks, Rong et al. designed a nano particle energy harvesting model \cite{7}. Although there have been some researches on SWIPT waveform design, segmentation coefficient optimization, SWIPT mechanism design, etc., there are few researches on SWIPT technology as an energy harvesting method for terahertz nano sensor networks. 
	
	Considering the limited energy of nano-nodes, we propose a SWIPT-based energy balance cluster network framework for WNSNs (EBCNF). The EBCNF network framework uses SWIPT technology to extend the network lifetime. As for the coefficient optimization problem in SWIPT, we transform the problem into a maximum-minimum problem for processing optimization. In addition, distance and energy are considered in the clustering process: the closer the nano-node is to the nano control node, the higher the probability of becoming a CH(cluster head node). The lower the energy of the CH and the closer the distance to the nanocontrol node, the smaller the cluster formed\cite{8}. This allows the CHs close to the nano control node to allocate a portion of energy to process data from other CHs.
	
	In Section 2, the network model, channel model and energy model of EBCNF are introduced. Next, the communication mechanism of EBCNF, cluster formation and update, and coefficient optimization are introduced in Section 3. Then, in Section 4, we demonstrate and analyze the superiority of EBCNF in terms of network survival time, data transmission success rate, throughput, and so on. Finally, in Section 5, we summarize the advantages of EBCNF and the directions that can be improved.

\section{System Model and Problem Formulation}
In WNSNs, clustering is generally used to divide and manage the network to reduce the pressure caused by increased network scale. For WNSNs, the ultimate goal is to transfer the data collected in the network to the macro network. Therefore, in addition to ordinary nano sensor nodes, there are also nano control nodes that connect macro networks and WNSNs. The role of the nano sensor node is mainly to collect data, package the data and send it to the nano control node. Data is transmitted from the nano sensor nodes to the CH, and then many CHs forward the information to the nano control node (NC).
	
	\subsection{Network Model}
	We use $G = \left( {V,E} \right)$ represent the network topology, where the set of nano-nodes is represented by$V = \left\{ {{v_1},{v_2}, \ldots ,{v_n}} \right\}$ , and $n = \left| V \right|$ is the toal number of nano-nodes. Figure 1 is a schematic diagram of the EBCNF we proposed. In the figure, there is a nano control node and multiple nano sensor nodes. NC can wirelessly charge all sensor nodes, and it can also be used as a sink node to collect information about each cluster. It should be noted that only NC can provide stable energy. NC regularly broadcasts terahertz waves, and all nano-nodes obtain energy from the terahertz waves. Several CHs are selected from the nano-nodes, a CH and surrounding nodes form a cluster, and the cluster size is determined according to the energy of the CH and the distance from the NC\cite{8}.

\begin{figure}[t]
		\centering
		\includegraphics[width=10cm]{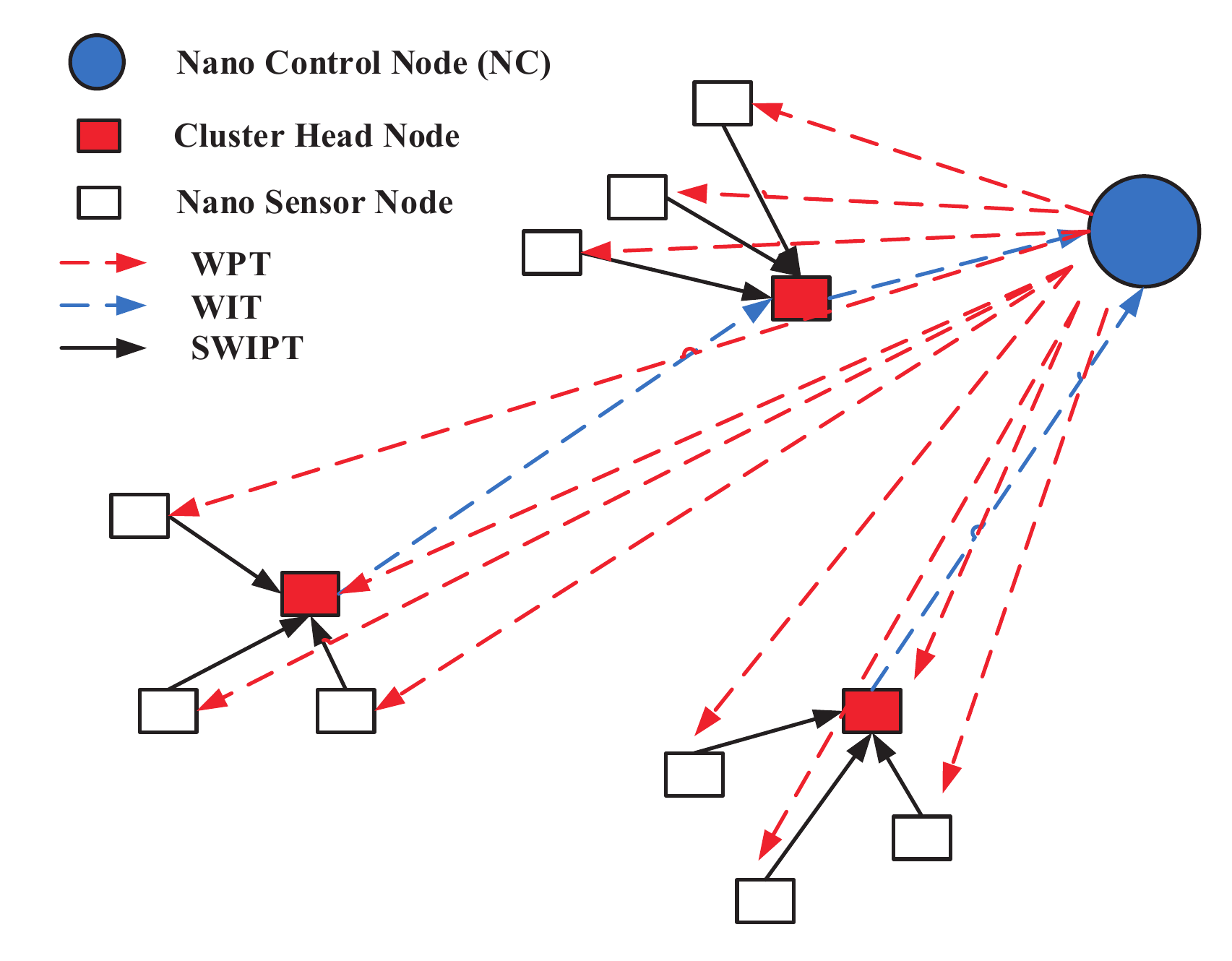}
		\caption{Schematic diagram of EBCNF framework.}
		\label{fig1}
\end{figure}
	
The establishment of the network model is based on the following assumptions:

	\begin{itemize}
	\item 	All nano-nodes in the network are in fixed positions, and the energy of NC is always sufficient.
	
	\item 	All nano-nodes sense volatile organic compounds, temperature and other information, and the environmental information sensed by each sensor node is transmitted through data packets of the same size.
	
	\item 	Nano-nodes can get the location information, remaining energy and channel quality of neighbor nodes through Hello messages, and all nano-nodes are identified by a unique ID.
	
	\item 	Nano-node processing data does not consume energy, while sending and receiving data consumes energy.
	
	\item 	Nano-node processing data does not consume energy, while sending and receiving data consumes energy.
	
	\item 	Node transmit power can be adjusted according to the specific transmission time slot length.
	
	\item 	The nano-node can perceive its own remaining energy value. Through WPT(Wireless Power Transfer) and SWIPT, the nanonode can collect energy from the environment through electromagnetic waves.
\end{itemize} 

	\subsection{Terahertz Channel Model}
	The path loss in the terahertz band can be expressed as the product of propagation loss and molecular absorption loss\cite{9}:
	\begin{equation}
		PL\left( {f,d} \right) = P{L_{spr}}\left( {f,d} \right) \times P{L_{abs}}\left( {f,d} \right)
	\end{equation}
where $P{L_{spr}}\left( {f,d} \right)$is the propagation loss, and $P{L_{abs}}\left( {f,d} \right)$ is the molecular absorption loss. $f$ is the transmission frequency and $d$ is the propagation distance\cite{10}.
The propagation loss can be expressed as:
	\begin{equation}
P{L_{spr}}\left( {f,d} \right) = {\left( {\frac{{4\pi fd}}{c}} \right)^2}
\end{equation}
where $c$ represents the speed of light in vacuum.
Under normal circumstances, when electromagnetic waves propagate in the medium, molecules will absorb part of the electromagnetic energy, causing molecular absorption loss. The magnitude of the absorption loss is related to the type of molecules present in the medium and the frequency of electromagnetic waves. The molecular absorption loss can be expressed as\cite{10}:
	\begin{equation}
	P{L_{abs}}\left( {f,d} \right) = {e^{k\left( f \right)d}}
\end{equation}
where $k\left( f \right)$ is the molecular absorption factor, which can be expressed as\cite{11}:
	\begin{equation}
	k\left( f \right) = \sum\limits_{i,g} {{k^{i,g}}\left( f \right)} 
\end{equation}
where ${{k^{i,g}}\left( f \right)}$ represents the absorption factor of the gas molecule $i$ in the medium $g$.
The channel capacity of a terahertz channel is equal to the sum of the capacities of the subchannels that make up the channel:
	\begin{equation}
C\left( d \right) = \sum\limits_i {\Delta f{{\log }_2}\left( {1 + \frac{{S\left( {{f_i}} \right)}}{{PL\left( {{f_i},d} \right)N\left( {{f_i},d} \right)}}} \right)} 
\end{equation}
where ${{f_i}}$ is the center frequency, ${\Delta f}$ is the bandwidth of the subchannels, ${S\left( {{f_i}} \right)}$ is PSD(power spectral density) of the transmitted signal, and ${N\left( {{f_i},d} \right)}$ is PSD of the noise in the channel\cite{12}.
Molecular absorption noise dominates THz channel noise sources, so ${N\left( {{f_i},d} \right)}$ can be expressed as PSD of molecular absorption noise:
	\begin{equation}
N\left( {f,d} \right) = {K_B}{T_0}\left( {1 - {e^{ - k\left( f \right)d}}} \right)
\end{equation}
where ${K_B}$ is Boltzmann's constant, and ${T_0}$ is the reference temperature.
	\subsection{Energy Consumption Model}
	The energy consumption model can be expressed as the energy used by nano-nodes to transmit and receive data:
		\begin{equation}
{E_{tot - con}} = {E_{tr - con}} + {E_{re - con}}
	\end{equation}
where ${E_{tr - con}}$ represents the energy used to transmit data and ${E_{re - con}}$ represents the energy used to receive data.
According to\cite{13}, the energy required by the nanonode to transmit data can be expressed as:
		\begin{equation}
{E_{tr - con}} = k\Delta fS\left( f \right){T_{bit}}
\end{equation}
where $k$ represents the number of bits of a data packet, $\Delta f$ represents the bandwidth of the current node's transmission signal, $S\left( f \right)$ represents PSD of the transmission signal, and ${T_{bit}}$ represents the time required to transmit 1 bit of data.

${E_{re - con}}$ represents the energy used for receiving data packets. In Section 2.1, it has been assumed that the environmental information sensed by each sensor node is transmitted through data packets of the same size. so we assume ${E_{re - con}}$ to be a constant $\phi $. Then the total energy consumption of the nanonode can be expressed as:
		\begin{equation}
	{E_{tot - con}} = k\Delta fS\left( f \right){T_{bit}} + \phi 
\end{equation}
	\subsection{Energy Harvesting Model}
	The energy harvesting model is an actual non-linear energy harvesting model based on the logistic function proposed in\cite{13,14,15}. Compared with the linear model, this model can capture the nonlinear behavior of the energy harvesting process.
	In the nonlinear model, the energy collected in the $k$th symbol interval can be expressed as:
			\begin{equation}
	{E_{har}} = \frac{{T{P_s}\left[ {\psi \left( {{\rho _k}} \right) - \gamma } \right]}}{{1 - \gamma }}
	\end{equation}
where $\gamma $ is a constant to ensure zero input and zero output response, ${\psi \left( {{\rho _k}} \right)}$ is a logistic function, ${{P_s}}$ is the maximum power at which the energy harvesting circuit is saturated\cite{16}, ${{\rho _k}}$ is the ratio of energy harvesting, and $T$ is the symbol interval.
$\gamma $ can be expressed as:
			\begin{equation}
\gamma  = \frac{1}{{1 + {e^{AB}}}}
\end{equation}
$\psi \left( {{\rho _k}} \right)$ can be expressed as:
			\begin{equation}
	\psi \left( {{\rho _k}} \right) = \frac{1}{{1 + {e^{ - A\left( {{\rho _k}hP - B} \right)}}}}
\end{equation}
where $h$ is the channel gain, ${\left| h \right|^2} = 1/PL\left( {f,d} \right)$. $P$ is the transmitted power. $A$ reflects the non-linear charging rate and $B$ is related to the turn-on threshold of the energy harvesting circuit.

\section{EBCNF Network Framework}
	\subsection{Communication Mechanism}
	The media access layer uses a hybrid MAC access mechanism of Carrier Sense Multiple Access with Collision Avoidance (CSMA/CA) and Time Division Multiple Access (TDMA). According to the acquisition-transmission protocol\cite{17} , NC first transmits radio frequency energy with a length of ${T_{wet}}$ through wireless energy transmission (WET). During this time slot, all nano-nodes will collect energy for use in subsequent data transmission. In the data transmission stage, the NC allocates the time slot of the node in the cluster and the time slot of the cluster head according to the amount of data to be transmitted in the cluster, and forwards the data to each cluster\cite{10}. The frame structure design is shown in Figure 2.
	\begin{figure}[t]
		\centering
		\includegraphics[width=10cm]{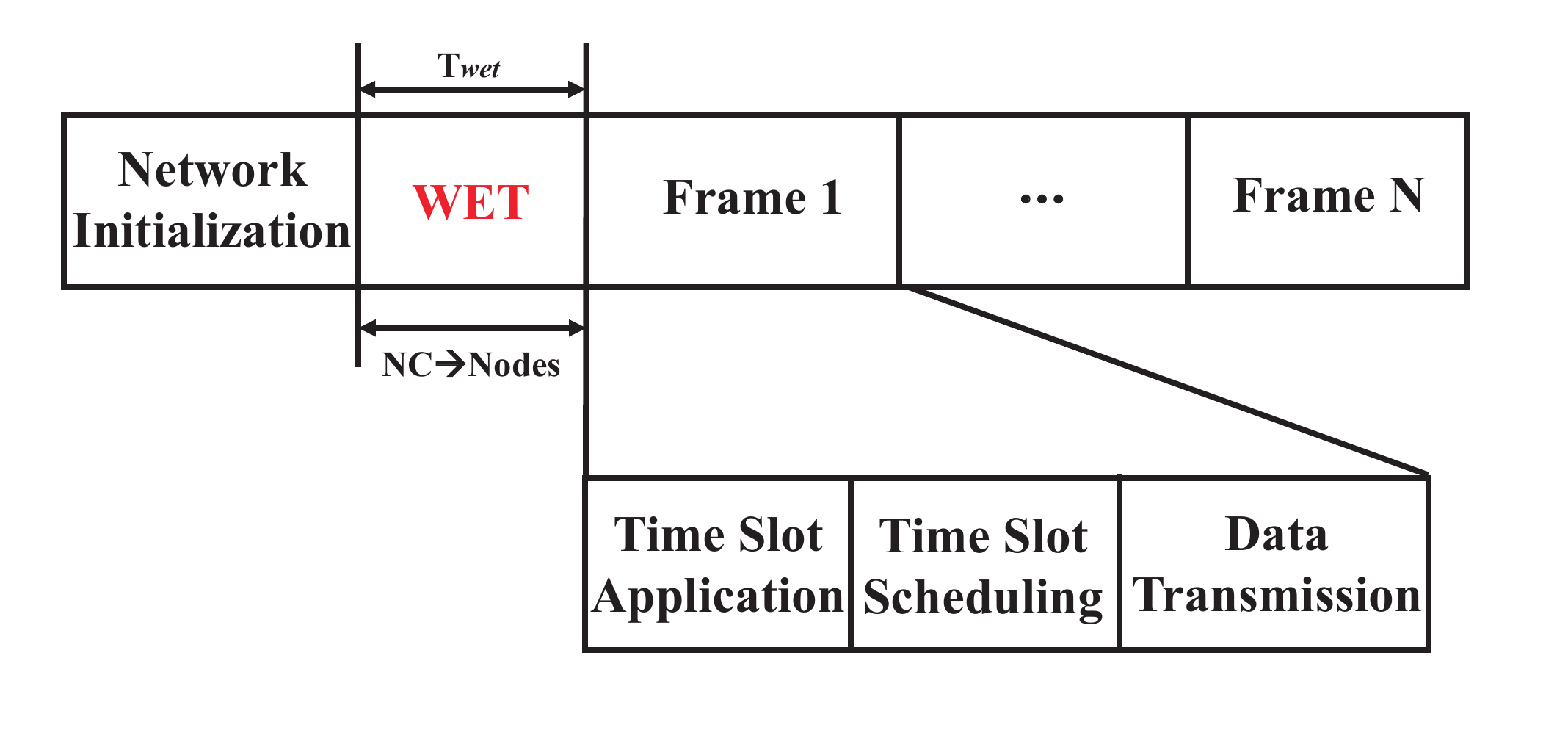}
		\caption{Frame structure.}
		\label{fig2}
	\end{figure}
	\begin{itemize}
	\item 		Time slot application phase: each cluster member node sends a time slot application packet RTS to the CH, and the time slot application packet contains the location information of the node and the amount of information transmitted. If a member node in the cluster has no data to send, the data amount is set to 0. CH receives the time slot request packet RTS from the node and replies with a CTS message. After the CH receives the time slot request packet from all member nodes, it calculates the total data transmission volume in the cluster and forwards the data packet to NC, also using the RTS/CTS mechanism\cite{10}.
	
	\item 		NC estimates the propagation time based on the RTS packets of all CHs and allocates the packet transmission sequence for inter-cluster nodes and intra-cluster nodes. The specific time schedule can be summarized into two stages:
	Stage 1: NC allocates transmission time slots for each cluster, and the length of the time slot is determined by the data volume parameter in the RTS packet that the current cluster head transmits to the NC.
	Stage 2: The cluster head allocates corresponding time slots for each nano-node in the cluster that needs to transmit data.
\end{itemize} 
  
  Figure 3 shows the time slot scheduling process in a cluster. For each cluster head, in its own time slot, the cluster head allocates time slots to active nano-nodes in the cluster. The nano-node uses the TS/PS mechanism in its own ${T_{sc}}$ time slot to transmit its remaining energy to the CH while transmitting information to CH. CH receives information and energy from nodes in the cluster according to the allocated time slots. Nodes in the cluster turn on the receiver in the ${T_{wet}}$ time slot for energy harvesting, turn on the transmitter during the ${T_{sc}}$ time slot for SWIPT transmission to the cluster head, turn off the transmitter and receiver during the rest of the time, and only perform the task of sensing until the beginning of the next frame. After that, the CH will fuse the data and forward it to the NC or the next hop CH.
  In the initial stage, except for CH, all nano-nodes are in data acquisition mode. After the time slot allocation is completed, the NC sends a wake-up code to activate the nano-node\cite{10}. The nano-node that has collected the data is activated to transmit data to the CH. After receiving all the data, the CH merges all the data and transmits it to the next hop nano-node during the allocated time slot.
  	\begin{figure}[t]
  	\centering
  	\includegraphics[width=10cm]{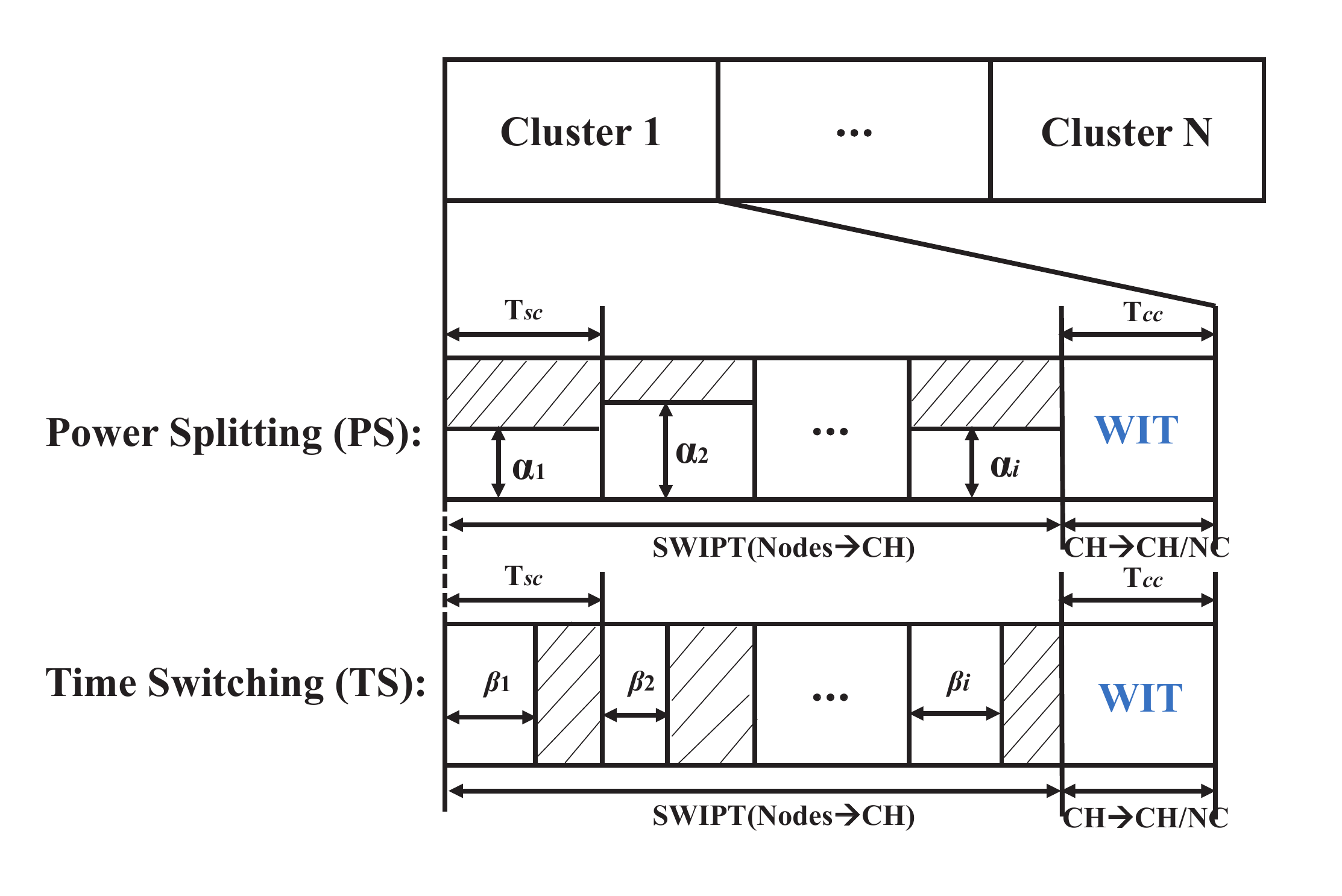}
  	\caption{Example of time slot scheduling in a cluster.}
  	\label{fig3}
  \end{figure}
	\subsection{Coefficient Optimization in SWIPT}
	Here we discuss two common mechanisms in SWIPT: PS and TS. The coefficient optimization in SWIPT is carried out on the basis of each cluster. In general, only one cluster can be considered. Assuming $m$ nano-nodes in the cluster, the $p$-th node is selected as CH, and the relationship between the $q$-th nanonode and CH is discussed.
	The energy obtained by the nano-node $q$ from the NC can be expressed as:
	\begin{equation}
		E_q^{har} = \frac{{P{T_{wet}}\left[ {\psi \left( {{\rho _k}} \right) - \gamma } \right]}}{{1 - \gamma }}
	\end{equation}
where $P$ is the signal power received by the nano-node. Since all electromagnetic waves are converted into energy in the process of NC transmitting energy to the nano-node, ${\rho _k} = 1$. 
According to (9), when the transmission frequency and data volume are determined, the energy consumed by the nano-node is fixed. Then the power that the nano-node can use for SWIPT can be expressed as:
	\begin{equation}
{P_q} = \frac{{{E_q} + E_q^{har} - E_q^{con}}}{{{T_{sc}}}}
\end{equation}
where ${{E_q}}$ represents the remaining energy of node $q$ and ${E_q^{con}}$ represents the energy consumed by node $q$.
Combining (5) and (14), the available transmission rate from nano-node $q$ to CH $p$ can be obtained as:
	\begin{equation}
	{R_{qp}} = \frac{1}{{{T_{sc}}}}{\log _2}\left( {1 + \frac{{{T_{sc}}{P_q}}}{{{{\left( {\frac{{4\pi fd}}{c}} \right)}^2}{e^{ - k\left( f \right)d}}{K_B}{T_0}\left( {1 - {e^{ - k\left( f \right)d}}} \right)}}} \right)
\end{equation}
For the cluster head $p$, if the SWIPT technology is not used to charge the cluster head, the CH can only use its energy and the energy obtained from the NC for data forwarding. Then the power of CH for data transmission in the WIT phase is similar to (14), which can be expressed as:
	\begin{equation}
	{P_p} = \frac{{{E_p} + E_p^{har} - E_p^{con}}}{{{T_{cc}}}}
\end{equation}
Without SWIPT, the transmission rate of CH $p$ can be expressed as:
	\begin{equation}
	{R_p} = \frac{1}{{{T_{cc}}}}{\log _2}\left( {1 + \frac{{{T_{cc}}{P_p}}}{{{{\left( {\frac{{4\pi fd}}{c}} \right)}^2}{e^{ - k\left( f \right)d}}{K_B}{T_0}\left( {1 - {e^{ - k\left( f \right)d}}} \right)}}} \right)
\end{equation}
According to \cite{18}, the rate of sensing data in the cluster should match the minimum rate of transmitting data, which can be expressed as:
	\begin{equation}
	{R_{no\_swipt}} = \min \left\{ {{R_{qp}},{R_p}} \right\}
\end{equation}
So the nano-nodes that can reach the transmission rate higher than the minimum rate of the transmission link only need to consume the energy required to meet the minimum rate, and then use SWIPT to transmit the remaining energy to the CH, and the CH uses the energy collected from each node to complete the subsequent data fusion processing and forwarding operations. Next, we will analyze the two SWIPT mechanisms, PS and TS.

For TS, nanonode $q$ uses ${\beta _q}{T_{sc}}$ time for information transmission, and the remaining $\left( {1 - {\beta _q}} \right){T_{sc}}$ time is used for energy transmission. Therefore, when using the TS mechanism, the achievable transmission rate from nano-node $q$ to CH is:
	\begin{equation}
	R_{qp}^{TS} = \frac{1}{{{\beta _q}{T_{sc}}}}{\log _2}\left( {1 + \frac{{{T_{sc}}{P_p}}}{{{{\left( {\frac{{4\pi f{d_{qp}}}}{c}} \right)}^2}{e^{ - k\left( f \right){d_{qp}}}}{K_B}{T_0}\left( {1 - {e^{ - k\left( f \right){d_{qp}}}}} \right)}}} \right)
\end{equation}

For TS, the energy that the cluster head $p$ can obtain from each nano-node is:

	\begin{equation}
E_p^{TS\_add} = \sum\limits_{\scriptstyle i = 0\hfill\atop
	\scriptstyle i \ne p\hfill}^m {\left( {1 - {\beta _i}} \right){P_i}{T_{sc}}} 
\end{equation}
where ${{P_i}}$ is the transmission power of nano-node $i$, and it can be calculated by (14).

For PS, the nano-node $q$ uses the power of ${\alpha _q}{P_q}$ for information transmission, and the remaining $\left( {1 - {\alpha _q}} \right){P_q}$ power is used for energy transmission. Therefore, when using the PS mechanism, the achievable transmission rate from nanonode $q$ to CH is:

	\begin{equation}
R_{qp}^{PS} = \frac{1}{{{T_{sc}}}}{\log _2}\left( {1 + \frac{{{\alpha _q}{T_{sc}}{P_p}}}{{{{\left( {\frac{{4\pi f{d_{qp}}}}{c}} \right)}^2}{e^{ - k\left( f \right){d_{qp}}}}{K_B}{T_0}\left( {1 - {e^{ - k\left( f \right){d_{qp}}}}} \right)}}} \right)
\end{equation}

For PS, the energy that the cluster head $p$ can obtain from each nano-node is:

	\begin{equation}
E_p^{PS\_add} = \sum\limits_{\scriptstyle i = 0\hfill\atop
	\scriptstyle i \ne p\hfill}^m {\left( {1 - {\alpha _i}} \right){P_i}{T_{sc}}} 
\end{equation}

It can be seen from (19) and (21) that the two equations are essentially the same. When using SWIPT, the transmission power of CH $p$ can be expressed as:

	\begin{equation}
	P_p^{swipt} = \frac{{{E_p} + E_p^{har} + E_p^{add} - E_p^{con}}}{{{T_{cc}}}}
\end{equation}
  where ${{E_p}}$ is the remaining energy of CH p, ${E_p^{har}}$ is the energy obtained by CH $p$ from the NC, ${E_p^{add}}$ is the energy provided by the intra-cluster nano-nodes to CH $p$ through SWIPT and ${E_p^{con}}$ is the consumed energy of CH $p$.

Therefore, the achievable transmission rate of CH $p$ can be expressed as:

\begin{equation}
R_p^{swipt} = \frac{1}{{{T_{cc}}}}{\log _2}\left( {1 + \frac{{{T_{cc}}P_p^{swipt}}}{{{{\left( {\frac{{4\pi f{d_p}}}{c}} \right)}^2}{e^{ - k\left( f \right){d_p}}}{K_B}{T_0}\left( {1 - {e^{ - k\left( f \right){d_p}}}} \right)}}} \right)
\end{equation}

Therefore, when the SWIPT technology is used, the required rate for the cluster to transmit data is:

\begin{equation}
{R_{swipt}} = \left\{ \begin{array}{l}
	\min \left\{ {{{\min }_{i \in m \cap i \ne p}}\left\{ {R_{ip}^{TS}} \right\},R_p^{swipt}} \right\},{\rm{ }}for{\rm{ TS}}\\
	\min \left\{ {{{\min }_{i \in m \cap i \ne p}}\left\{ {R_{ip}^{PS}} \right\},R_p^{swipt}} \right\},{\rm{ }}for{\rm{ TS}}
\end{array} \right.
\end{equation}

For a given cluster head $p$, the coefficient of TS/PS is optimized to maximize ${R_{swipt}}$. Under the PS mechanism, it can be expressed as:

\begin{equation}
{\alpha ^*} = argmax{R_{swipt}} = {\mathop{\rm argmax}\nolimits} \min \left\{ {{{\min }_{i \in m \cap i \ne p}}\left\{ {R_{ip}^{PS}} \right\},R_p^{swipt}} \right\}
\end{equation}

Under the TS mechanism, it can be specifically expressed as:

\begin{equation}
{\beta ^*} = argmax{R_{swipt}} = {\mathop{\rm argmax}\nolimits} \min \left\{ {{{\min }_{i \in m \cap i \ne p}}\left\{ {R_{ip}^{TS}} \right\},R_p^{swipt}} \right\}
\end{equation}

According to (21) and (24), when the partition coefficient $\alpha $ of PS increases, ${R_{ip}^{PS}}$ increases, while ${R_p^{swipt}}$ decreases, and vice versa. Therefore, ${{{\min }_{i \in m \cap i \ne p}}\left\{ {R_{ip}^{PS}} \right\}}$ and ${R_p^{swipt}}$ need a compromise. The degree of this compromise is determined by the coefficient $\alpha $. It is similar for TS, and the degree of compromise is determined by the coefficient $\beta $.

Therefore, the problem is transformed into a maximum-minimum problem. We refer \cite{19} to optimize the coefficients of the problem under the premise of meeting the minimum transmission rate in the cluster. The coefficient optimization process is shown in Algorithm 1.

	\begin{algorithm}[t]
	{\footnotesize{
			\caption{Iterative Algorithm for SWIPT Optimal Coefficient}
			\SetAlgoLined 
			 $p$ = Cluster Header Node\;
			$i$ = Nano Sensor Node, $i \in m \cap i \ne p$\;
			$m$ = Set of Alive Nodes in the Cluster\;
			{{Initialize}} ${\rm{R}}_p^{res} \leftarrow {\min _{_{i \in m \cap i \ne p}}}\left\{ {{R_{pi}}} \right\}$\; 
			\While{$R_p^{swipt} < R_p^{res}$}{
				\uIf{PS is used for SWIPT}{
				${\alpha _i} \leftarrow \left( {{2^{R_p^{res}{T_{SC}}}} - 1} \right)\frac{{{K_B}{T_0}\left( {1 - {e^{ - k\left( f \right)d}}} \right)}}{{{T_{SC}}{P_q}}}{\left( {\frac{{4\pi fd}}{c}} \right)^2}{e^{ - k\left( f \right)d}}$ from (21)\;
				}
			\ElseIf{TS is used for SWIPT}{
			${\beta _i} \leftarrow \frac{1}{{R_p^{res}{T_{SC}}}}{\log _2}\left( {1 + \frac{{{T_{SC}}{P_q}}}{{{{\left( {\frac{{4\pi f{d_{qp}}}}{c}} \right)}^2}{e^{ - k\left( f \right){d_{qp}}}}{K_B}{T_0}\left( {1 - {e^{ - k\left( f \right){d_{qp}}}}} \right)}}} \right)$ from (19)\;
		}
			Update $E_p^{add}$ according to (20)/(22)\;
			Update $R_p^{swipt}$ according to (24)\;
			$R_p^{res} \leftarrow \frac{{R_p^{swipt} + R_p^{res}}}{2}$\;

}
			\textbf{return} ${\alpha _i}$ or ${\beta _i}$.}}
\end{algorithm}

\subsection{Cluster Formation and Cluster Head Update}
We propose an energy balance algorithm for intra-cluster and inter-cluster nodes (EBACC). The nanonetwork is divided into clusters of different sizes when considering the update of CH in the clusters. The closer to the NC, the smaller the number of cluster nodes, ensuring more uniform energy consumption among cluster heads\cite{20}. The clustering diagram of this algorithm is shown in Figure 4.

	\begin{figure}[t]
	\centering
	\includegraphics[width=9cm]{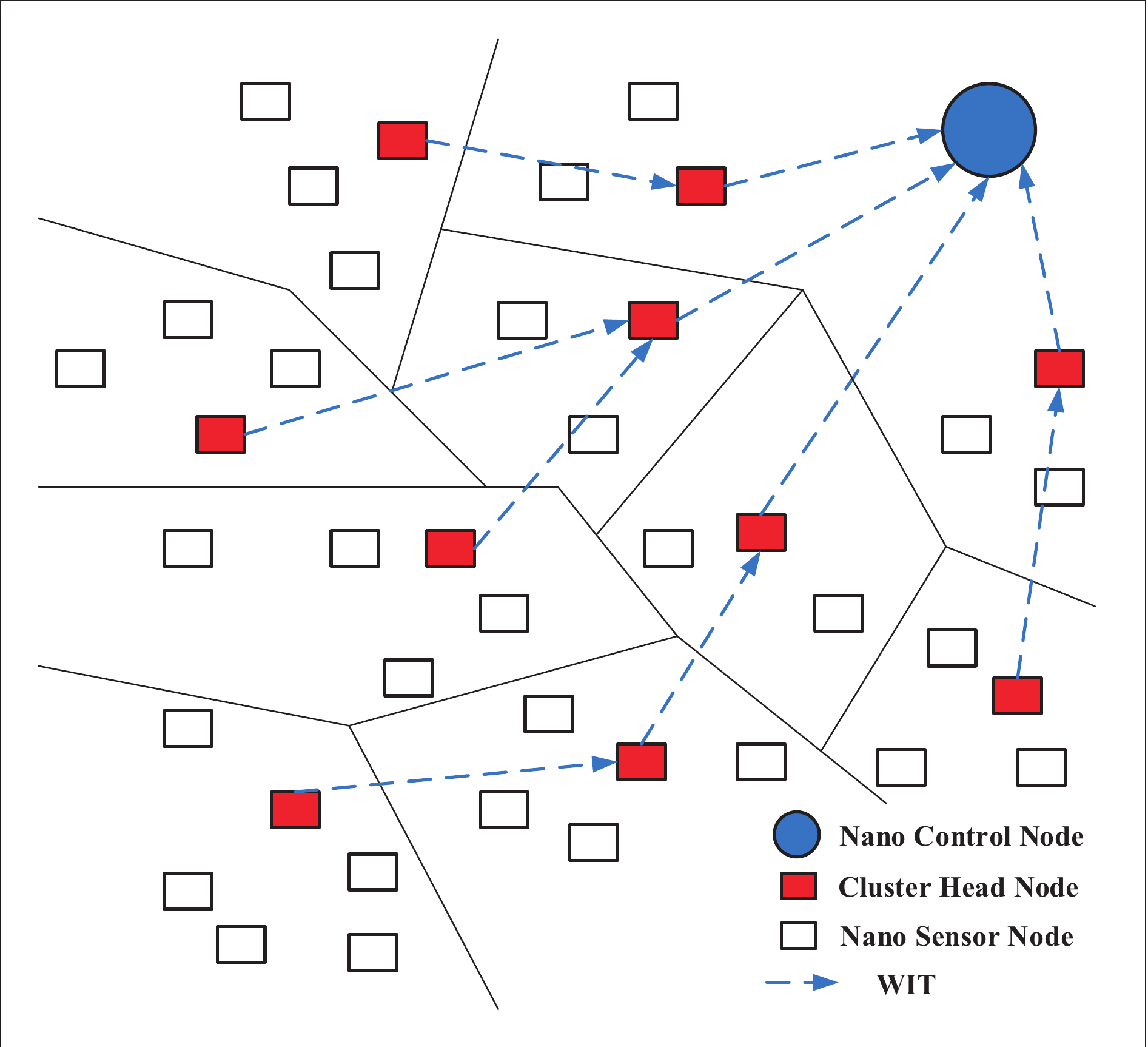}
	\caption{Schematic diagram of EBACC clustering.}
	\label{fig4}
\end{figure}

When using the EBACC algorithm for clustering, first, it is necessary to randomly select several nano-nodes in the network as candidate CHs to participate in CH competition. For ordinary nano-nodes, there is a certain probabi-lity to become a candidate CH. For nano-nodes that fail to become a candidate CH, they will stay in a dormant state and will not be awakened until the end of the cluster head selection phase. Since the range of clusters near the NC is smaller, that is to say, more clusters need to be generated near the NC. Therefore, in the cluster head selection stage, the probability that the node close to the NC becomes the candidate CH is increased. The specific threshold calculation equation for the selection of the candidate CH is:

\begin{equation}
T\left( i \right) = \frac{p}{{1 - p\left( {r\bmod \left( {\frac{1}{p}} \right)} \right)}} \cdot \frac{{{d_{\max }} - d\left( {a,NC} \right)}}{{{d_{\max }} - {d_{\min }}}}
\end{equation}

where $p$ is the percentage of node ${v_i}$ that becomes CH, $r$ is the current round, ${{d_{\max }}}$ represents the maximum distance from the nano-node to the NC, ${{d_{\min }}}$ represents the minimum distance from the nanonode to the NC, and ${d\left( {a,NC} \right)}$ represents the distance from nanonode a to NC.

In the cluster establishment stage, the nano-node ${v_i}$ will generate a random number between 0-1, if the generated random number is less than the threshold $T\left( i \right)$, the node becomes a candidate CH\cite{8}.

Secondly, for each candidate CH, it will have its own competition distance $R$. The larger the $R$ value, the larger the cluster size. Within the scope of a cluster, only one CH is allowed, that is, if the nano-node $a$ becomes a CH, then there will no other CH in the competition range $R$ of $a$. Therefore, measures need to be taken to control the size of the cluster. CHs closer to the NC should maintain smaller clusters. At the same time, since the range of clusters near the NC is smaller, more clusters should be formed near the NC. In addition to the distance from the NC, if the energy of the current candidate CH is larger, a larger cluster can be maintained. That is, the competition range of a candidate CH increases with its distance from the NC and its own remaining energy. The competition range ${R_a}$ of the candidate CH $a$ is expressed as a linear function of the distance and energy from the NC:

\begin{equation}
{R_a} = \left( {1 - a\frac{{{d_{\max }} - d\left( {a,NC} \right)}}{{{d_{\max }} - {d_{\min }}}} - b\frac{{{e_{\max }} - {e_i}}}{{{e_{\max }}}}} \right){R_0}
\end{equation}

where ${R_0}$ is the pre-defined maximum competition range, $a$ and $b$ are constant coefficients between $0 - 1$, ${{e_{\max }}}$ represents the maximum energy of the nano-node, and ${{e_i}}$ represents the remaining energy of the nano-node ${v_i}$.

Each candidate CH also needs to maintain a set ${S_{CH}}$ of its neighbor candidate CHs. Specifically, if there is another candidate CH b in the competition range ${R_a}$ of the candidate CH $a$, then $b$ is an adjacent candidate CH of $a$ and $b$ needs to be added to the ${S_{CH}}$ of $a$. After the construction of the set ${S_{CH}}$ of $a$ is completed, $a$ needs to compare the residual energy (RE) with the candidate CHs in its ${S_{CH}}$. If $a$ finds that its remaining energy is greater than the adjacent candidate CH, then it sends ${\rm{GIVE\_UP\_MSG}}$; if $a$ receives ${\rm{GIVE\_UP\_MSG}}$ sent by other nodes, it sends ${\rm{NOMORE\_CH\_MSG}}$ and gives up the cluster head competition to become a normal nano-node; if $a$ receives ${\rm{NOMORE\_CH\_MSG}}$ from other candidate CHs, $a$ deletes it from its ${S_{CH}}$.

After the CH selection is completed, the sleeping nano-nodes will be awakened, and each CH will broadcast a ${\rm{CH\_ADV\_MSG}}$ through the network field. Each nano-node selects the CH closest to itself and has the highest received signal strength, and then sends a ${\rm{JOIN\_CLUSTER\_MSG}}$ to notify the CH to become a node in its cluster.
Taking the candidate CH a as an example, the process of the EBACC algorithm is shown in Algorithm 2.

	\begin{algorithm}[t]
	{\footnotesize{
			\caption{EBACC}
			\SetAlgoLined 
			A cluster head competitive process for node $a$\;
			$\rho  \leftarrow Rand\left( {0,1} \right)$\;
			\If{$\rho  < T$}{
			$beCandidateHead \leftarrow True$\;	
		}
			\eIf{$beCandidateHead = True$}{
				broadcast COMPETE\_HEAD\_MSG($a$.ID, $a$.ENERGY)\;}{
				\textbf{exit}}
			\If{$a$ is a candidate node which receives a COMPETE\_HEAD\_MSG from
			 $b$ and  $d\left( {a,b} \right) < \max \left( {{R_a},{R_b}} \right)$}{
				add $b \to a.{S_{CH}}$\;	
			}
			\While{the time slot for cluster head competition has not 
				expired}{
							\If{$a.{\rm{ENERGY}} > b.{\rm{ENERGY}}$ and 
							$b \in a.{S_{CH}}$}{
					broadcast E(GIVE\_UP\_MESSAGE$a$.ID) and \textbf{exit}\;	
				}
							\If{$a$ receives GIVE\_UP\_MESSAGE from $b$ and $b \in a.{S_{CH}}$ }{
								broadcast NOMORE\_CH\_MSG($a$.ID) and \textbf{exit}\;	
							}
							\If{$a$ receives NOMORE\_CH\_MSG from $b$ and $b \in a.{S_{CH}}$ }{
								remove $b$ from $a.{S_{CH}}$\;	
							}
			}
		}}
\end{algorithm}

\section{Simulation Analysis}
We use the routing protocol in \cite{21} as the inter-cluster routing method, analyze EBCNF from the perspective of using the SWIPT mechanism and not using the SWIPT mechanism, and secondly needs to verify the effectiveness of the clustering algorithm. We compare and simulate the EBACC algorithm and the classic clustering protocol LEACH \cite{22}. The following four schemes are specifically compared:
	\begin{itemize}
	\item 	LEACH: CH is selected randomly according to a random threshold algorithm, and SWIPT is not used.
	
	\item 	EBACC: Use the EBACC clustering algorithm to create clusters, and use the method of \cite{21} for routing and forwarding between clusters, without using SWIPT.
	
	\item 	EBCNF based on PS mechanism: adopt the EBACC clustering and cluster head selection algorithm and the PS-based SWIPT.
	
	\item 	EBCNF based on TS mechanism: adopt the EBACC clustering and cluster head selection algorithm and the TS-based SWIPT.
\end{itemize} 
	\subsection{Evaluation Index}
		\begin{itemize}
		\item 	Network lifetime
		
		The rounds of iteration when the nodes in the network start to die are adopted to represent the lifetime of the network. Specifically, it can be expressed as:
		\begin{equation}
LT = \min \left\{ {r|{E_v}\left( r \right) \le \delta } \right\},v \in V
		\end{equation}
	where the network lifetime is represented by $LT$, $v$ is the nano-node in WNSN, $V$ is the set of nodes. After $r$ rounds of iterations, if the node energy ${{E_v}\left( r \right)}$ is less than $\delta $, the node is called a dead node.
		
		\item 	Average remaining energy
		
		The average remaining energy can be expressed as the ratio of the energy of all nodes in the current network to the energy of all nodes in the initial network\cite{10}:
		\begin{equation}
\overline {{E_{res}}}  = \frac{{\sum\limits_{i = 1}^n {{E_{res\_i}}} }}{{n{E_{init}}}}
		\end{equation}
		Where the network has $n$ nano-nodes, ${{E_{res\_i}}}$ is the remaining energy of the nano-node, and ${{E_{init}}}$ is the initial energy of the nanonode\cite{10}.
		
		\item 	Transmission success rate
		
		The transmission success rate is expressed as the ratio of the number of packets successfully sent to the NC to the total number:
		\begin{equation}
{R_{suc}} = \frac{{{D_r}}}{{{D_t}}}
\end{equation}
where ${{D_r}}$ represents the number of packets successfully sent to the NC, ${{D_t}}$ represents the total number of data packets sent by nodes in the network.

		\item 	Average throughput
		The average throughput can be expressed as the number of data packet bits successfully received by the NC per unit time:
		\begin{equation}
	\overline S  = \frac{{{D_{r\_T}} \cdot {N_{bit}}}}{T}
\end{equation}	
		where $\overline S $represents the average throughput, ${{D_{r\_T}}}$ represents the number of data packets successfully received by NC within $T$ time, and ${{N_{bit}}}$ represents the number of bits of a single data packet.	
		
		\item 	Control overhead
		
		The control overhead is defined as the total number of bytes of data packets used for operations such as network clustering and time slot request in the network. In the actual simulation, the network overhead is measured by controlling the cost ratio, that is, the proportion between the number of bytes of the control overhead of the nanometer node and the number of bytes of all packets in the network\cite{23}.
	\end{itemize} 

	\subsection{Simulation Parameter}
	The simulation scene is a two-dimensional square plane of 10mm×10mm. Nanonodes are randomly distributed in this plane. There is one and only one NC located on the right side of the square area, which is balanced with the center line of the square area, and is 1mm from the right side of the square (11mm, 5mm), as shown in Figure 5.
	\begin{figure}[t]
	\centering
	\includegraphics[width=15cm]{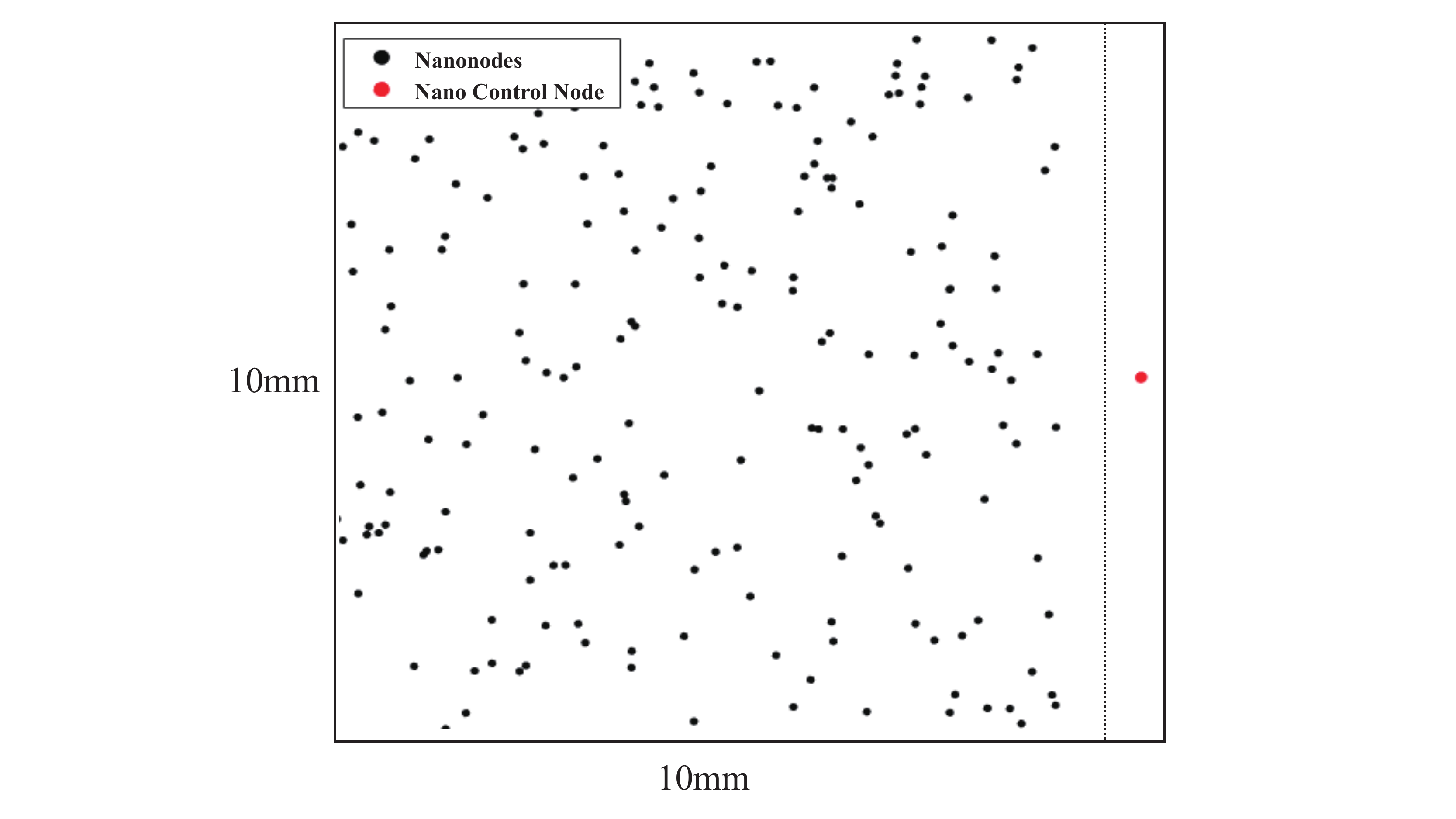}
	\caption{Network nodes deployment.}
	\label{fig5}
\end{figure}

In order to simplify the calculation, the terahertz frequency band of 0.5-1.5 THz is used, and $\Delta f$ is set to 0.01 THz \cite{4}. Considering the channel environment where the water molecule content is 10\%, the value of the absorption factor in (3) is 0.25 \cite{11}. The $\phi $ in (9) is set to 22nJ \cite{13}. The value of ${\rho _k}$ in (10) is 1 when the NC charges each nano-node. When the nano-node charges the cluster head, ${\rho _k} = \beta $ in the TS mechanism and ${\rho _k} = \alpha $ in the PS mechanism. In (11), $A$ is 6400 and $B$ is 0.003 \cite{14}. ${R_0}$ in (29) is set to 2mm\cite{24}, and parameters $a$ and $b$ are both set to 0.2. The remaining energy threshold $\delta $ in (30) is set to $1.4 \times {10^{ - 13}}J$. 

	\subsection{EBCNF Framework Simulation}
	\begin{itemize}
		\item 	Network lifetime
		
We define the survival period as the first round of dead nodes in the network. Figure 6 and Figure 7 are simulations of the network lifetime and the number of dead nodes in the network, respectively. Figure 6 shows the average number of rounds in which each protocol first had a dead node after running the network 2-20 times (with parameters re-initialized before each network run). It can be seen from Figure 6 that the first dead node in the LEACH protocol appears in the earliest round, which means that the network lifetime of the LEACH protocol is the shortest. It can be seen that the LEACH protocol cannot balance the nano-nodes well. The energy consumption of certain nodes is too high, and the remaining energy is lower than the energy threshold. The EBACC protocol balances the energy of nano-nodes through uneven clustering, which delays the appearance of the first dead node by about 200-300 rounds. In addition, PS-EBCNF and TS-EBCNF use the SWIPT mechanism to charge the nano-nodes in the network. The nano-nodes can also provide a certain degree of energy replenishment while consuming energy. The SWIPT charging mechanism makes the life of the network increase about 100-300 rounds on the basis of EBACC.
Figure 7 analyzes the lifetime of the network by counting the number of dead nodes in the network operation. Because both PS-EBCNF and TS-EBCNF adopt the SWIPT mechanism to replenish the energy of the nano-nodes in the network, thus the time when the nodes in the network start to die is delayed, and due to the use of the charging mechanism, The nano-nodes will not all die, but after 1100 rounds, a certain number of nodes survive stably. Since LEACH and EBACC have no charging mechanism, eventually all nodes in the network will die. However, the EBACC protocol considers the remaining energy of the CH and the distance to NC when determining the size of the cluster, and the LEACH protocol adopts a strategy of randomly selecting the CH and does not consider the size of the cluster. Therefore, the EBACC protocol effectively prolongs the survival time of nano-nodes compared to the LEACH protocol.

			\begin{figure}[h]
			\centering
			\includegraphics[width=8cm]{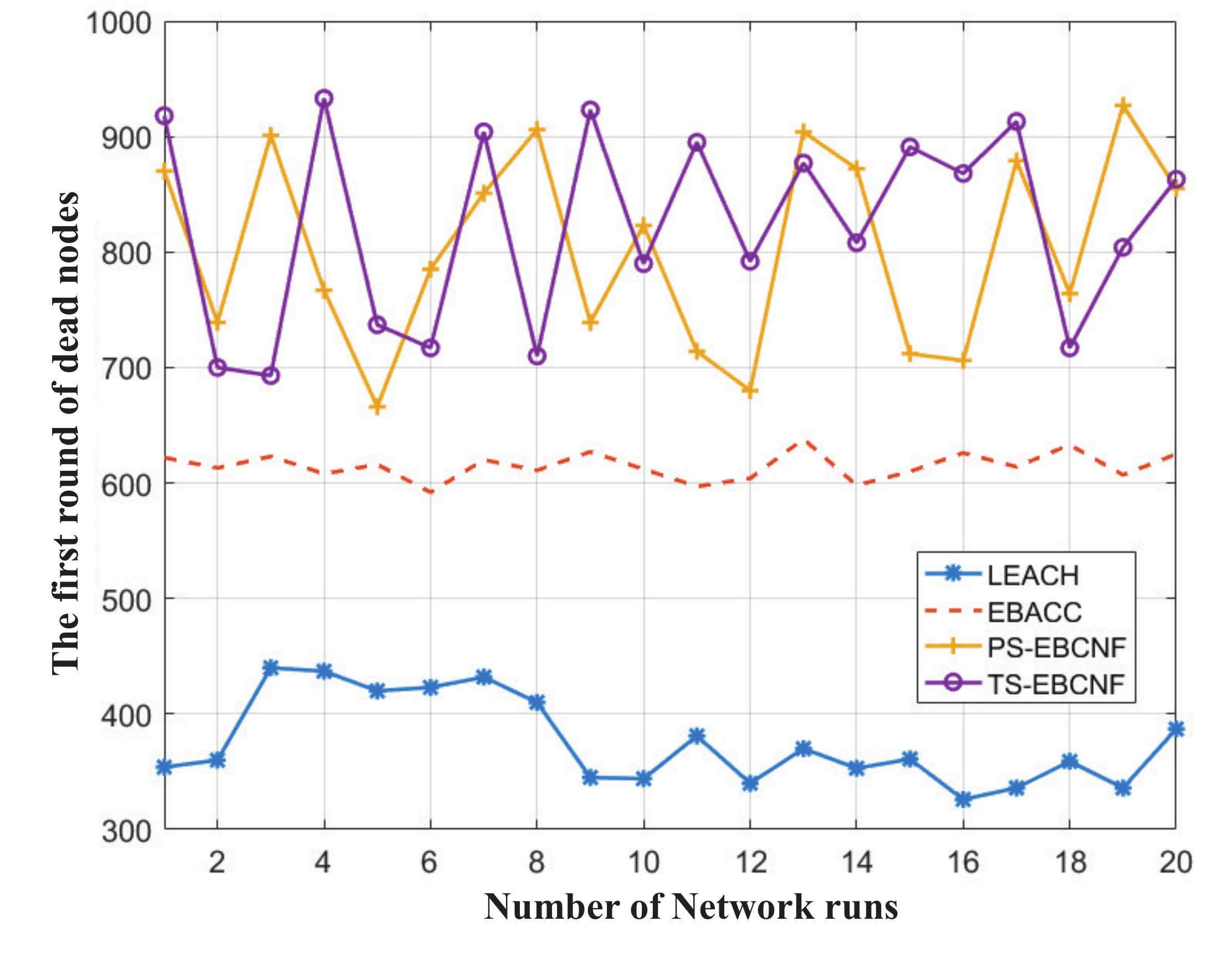}
			\caption{Comparison of network lifetime.}
			\label{fig6}
		\end{figure}
	\FloatBarrier
	
				\begin{figure}[h]
		\centering
		\includegraphics[width=8cm]{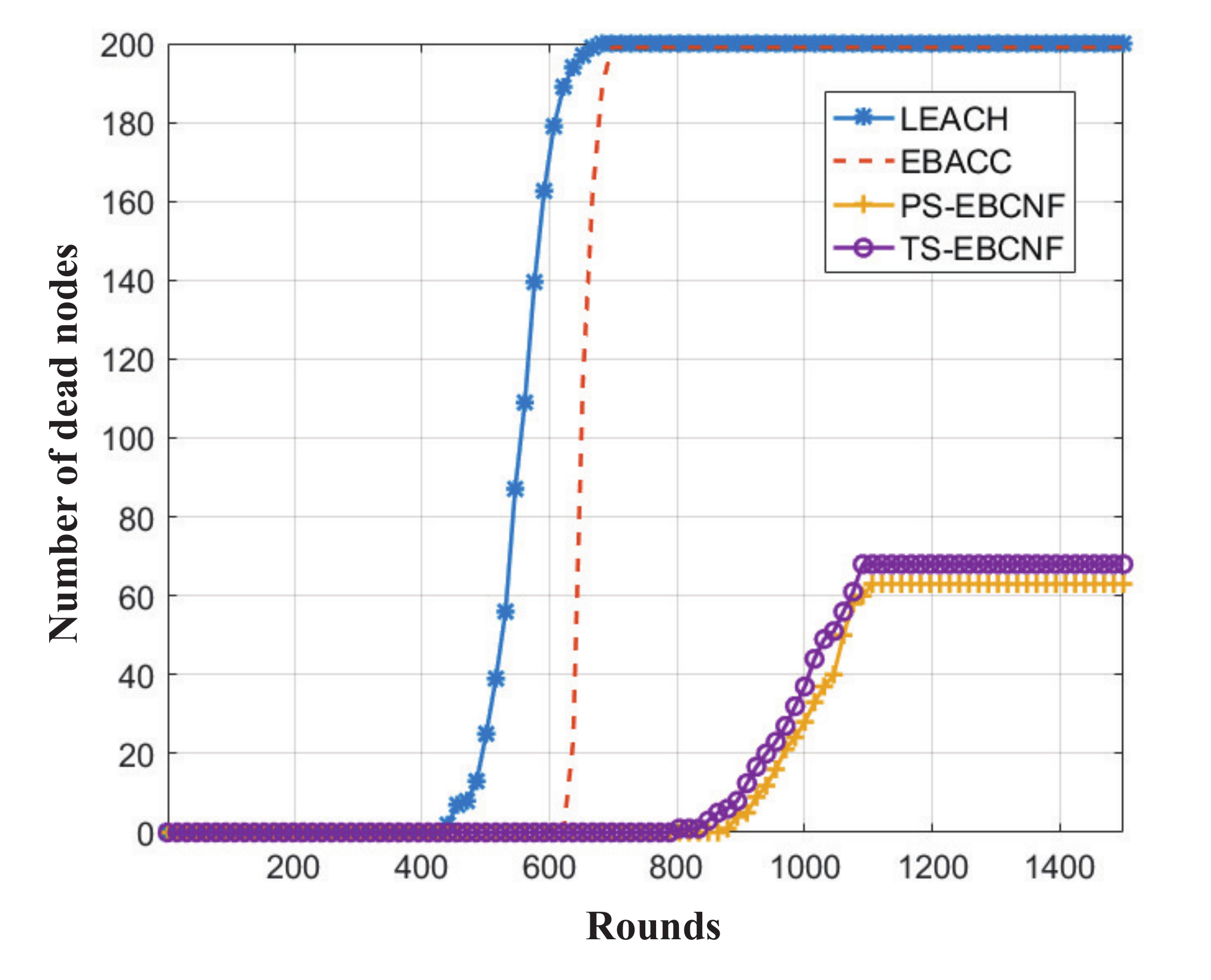}
		\caption{Comparison of the number of dead nodes.}
		\label{fig7}
	\end{figure}
\FloatBarrier

		\item 	Average remaining energy
		
Figure 8 shows the relationship between the average remaining energy difference of LEACH, EBACC, PS-EBCNF, TS-EBCNF and the rounds of network running 
time. It can be seen from the figure that compared with 
the LEACH protocol, the EBACC protocol has more 
residual energy, which indicates that the EBACC 
algorithm can make the energy consumption of nano-nodes more uniform through the measure of uneven 
clustering. PS-EBCNF and TS-EBCNF use the SWIPT 
mechanism and on the basis of the EBACC protocol to 
delay the death time of nano-nodes in the network through 
energy harvesting. Due to the charging mechanism, 
redundant nodes are deployed in the network, so that the 
nodes that have lost energy can regain sufficient energy 
through the charging of subsequent frames, and finally the 
life of the network tends to infinity.

				\begin{figure}[t]
	\centering
	\includegraphics[width=8cm]{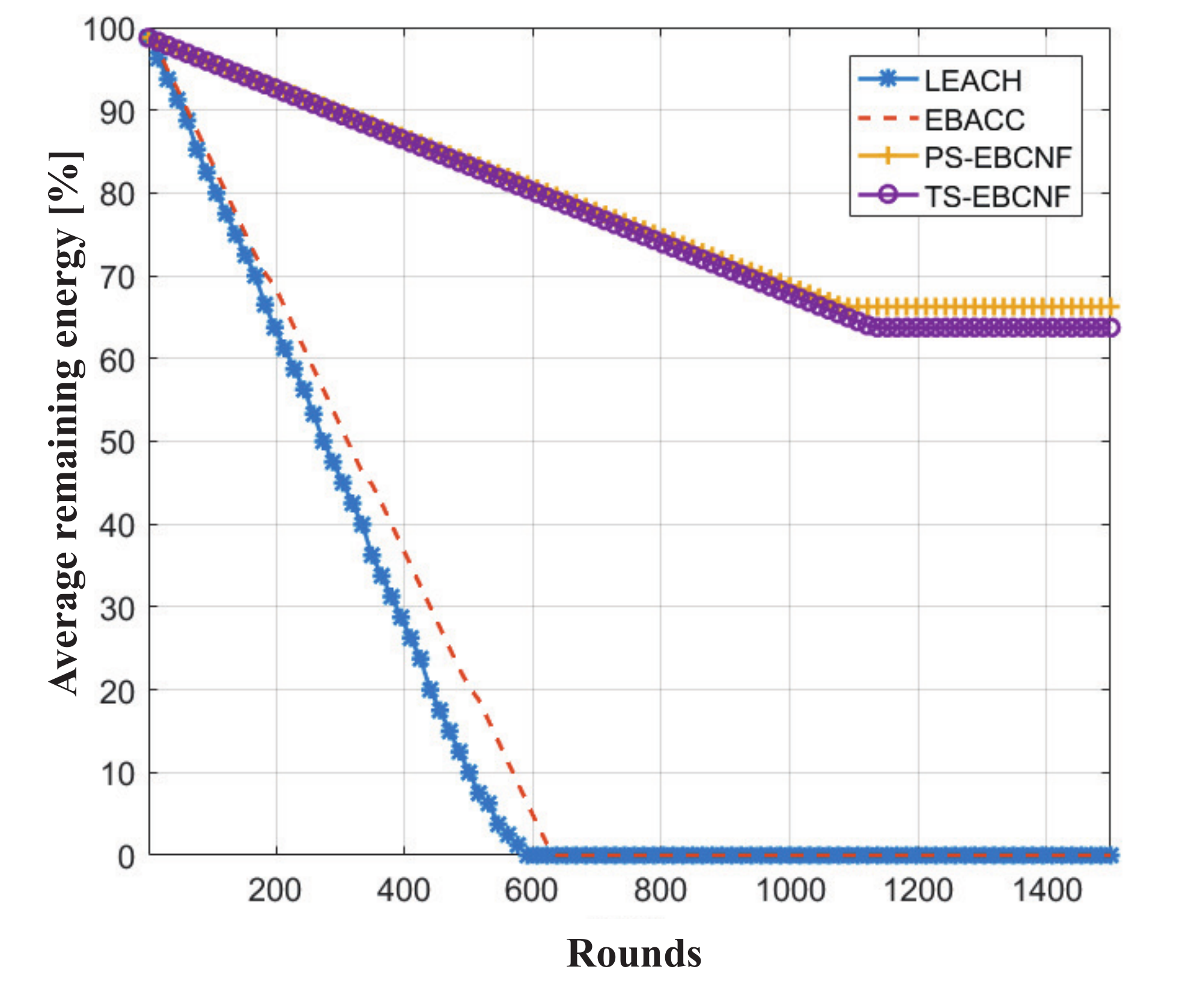}
	\caption{Comparison of average remaining energy.}
	\label{fig8}
\end{figure}
\FloatBarrier

		\item 	Transmission success rate
		
		Figure 9 shows the relationship between the packet 
		transmission success rate difference of LEACH, EBACC, 
		PS-EBCNF, TS-EBCNF and the interval size when 
		generating packets. As the interval size increases, the 
		transmission success rate also gradually increases. Because when the interval size is small, the nano-node
		needs to consume more energy for data transmission at the 
		same time, which accelerates the death of the node and the 
		burden on the network. For TS-EBCNF and PS-EBCNF, 
		when the interval size is small, the rate of energy 
		consumption exceeds the rate of energy absorption. Due 
		to insufficient energy, the nano-node cannot forward data 
		packets and packet loss occurs. When the interval size is 
		greater than 0.06s, node energy absorption and 
		consumption rate balance, so transmission success rate 
		increases. In terms of time, compared with LEACH 
		protocol, the case of EBACC protocol that cannot be 
		forwarded due to insufficient node energy is less than that 
		of LEACH protocol. Therefore, from the perspective of 
		the transmission success rate, the effect of the EBACC
		protocol is better than that of the LEACH protocol.
		
						\begin{figure}[h]
			\centering
			\includegraphics[width=8cm]{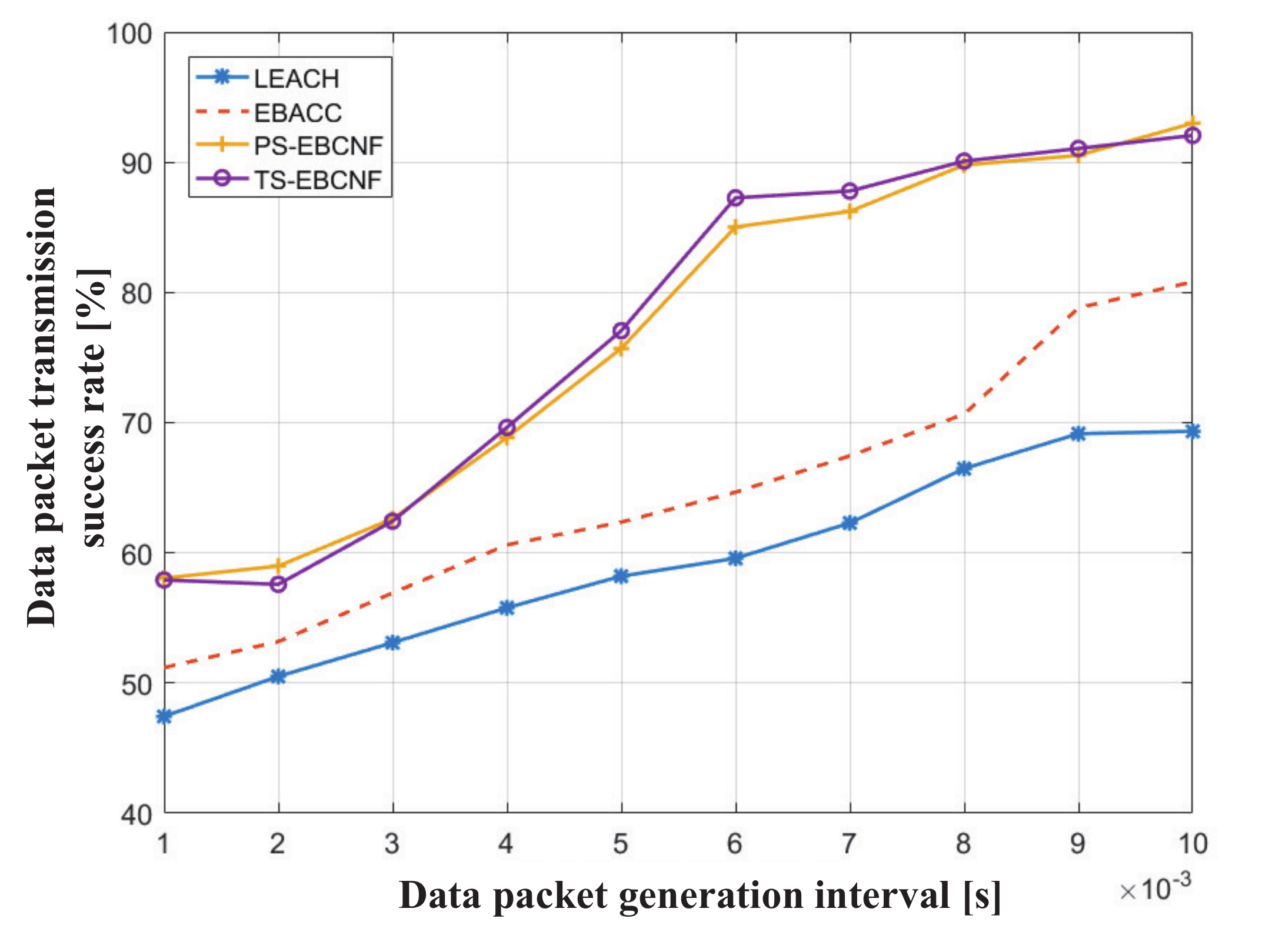}
			\caption{Comparison of data packet transmission success rate.}
			\label{fig9}
		\end{figure}
	\FloatBarrier

		\item 	Average throughput
		
		Figure 10 shows the relationship between the average 
		throughput difference of LEACH, EBACC, PS-EBCNF, 
		TS-EBCNF and the interval size when generating packets.
		When the interval size is constant, the more successful 
		packets are transmitted in the network, the higher the 
		throughput. From the previous analysis of the transmission success rate, the transmission success rate of the four protocols is ${\rm{TS - EBCNF}} \approx {\rm{PS - EBCNF > EBACC > LEACH}}$, so in terms of average throughput, TS-EBCNF and PS-EBCNF are better than EBACC, and EBACC is better than LEACH.
		
								\begin{figure}[h]
			\centering
			\includegraphics[width=8cm]{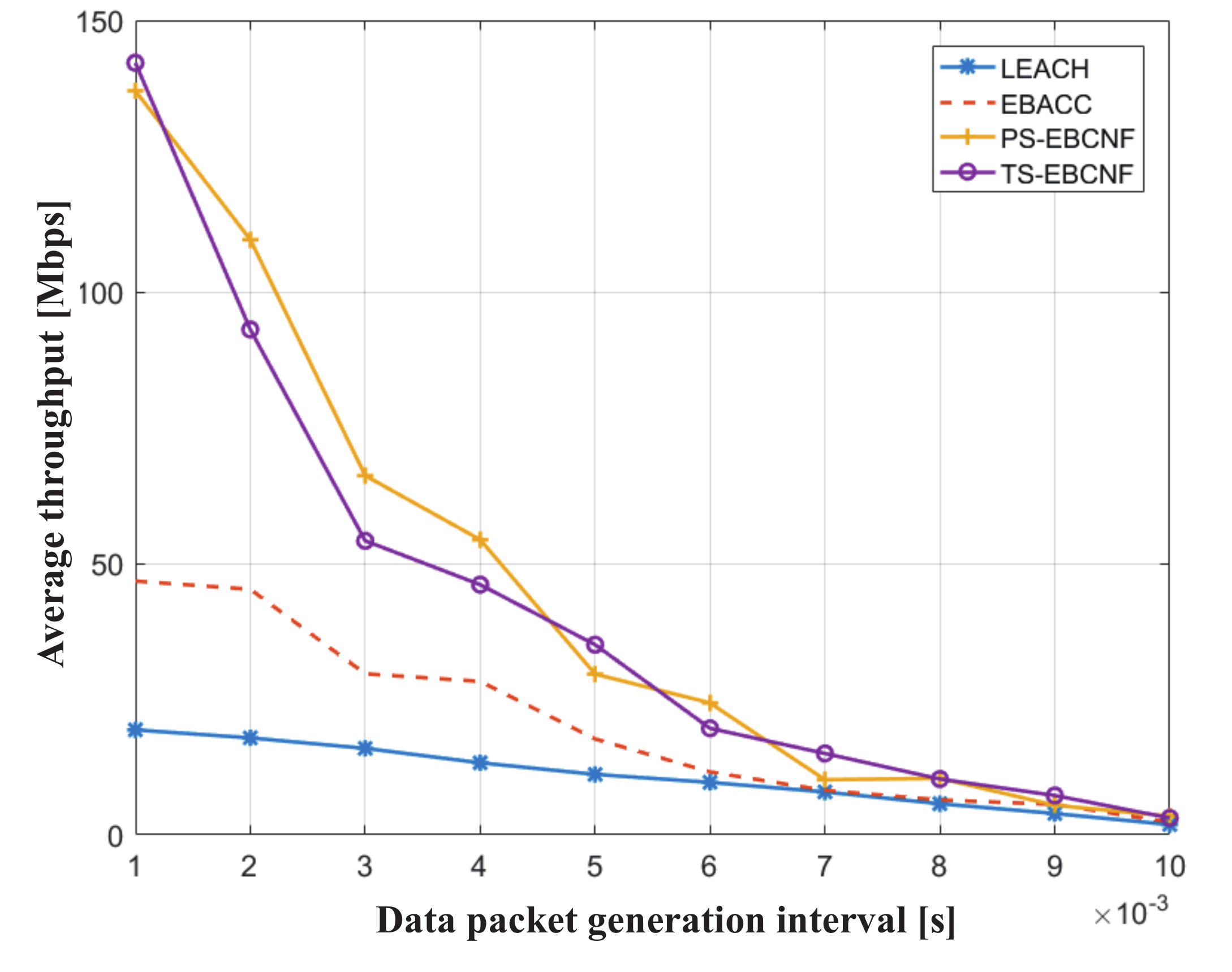}
			\caption{Comparison of average throughput.}
			\label{fig10}
		\end{figure}
	\FloatBarrier

		\item 	Control overhead
		
Figure 11 is a comparison of the control overhead of the four protocols. The control overhead of the EBCNF framework with the addition of uneven clustering and SWIPT mechanism is always the highest. This is because in addition to clustering, the EBCNF framework also requires the use of the SWIPT mechanism designed to 
improve the lifetime of the network to charge nodes, so 
the control overhead is slightly larger. However, through 
the SWIPT mechanism, the success rate of data packet 
transmission has been improved, so this part of the 
redundant control overhead is tolerable.

								\begin{figure}[t]
	\centering
	\includegraphics[width=8cm]{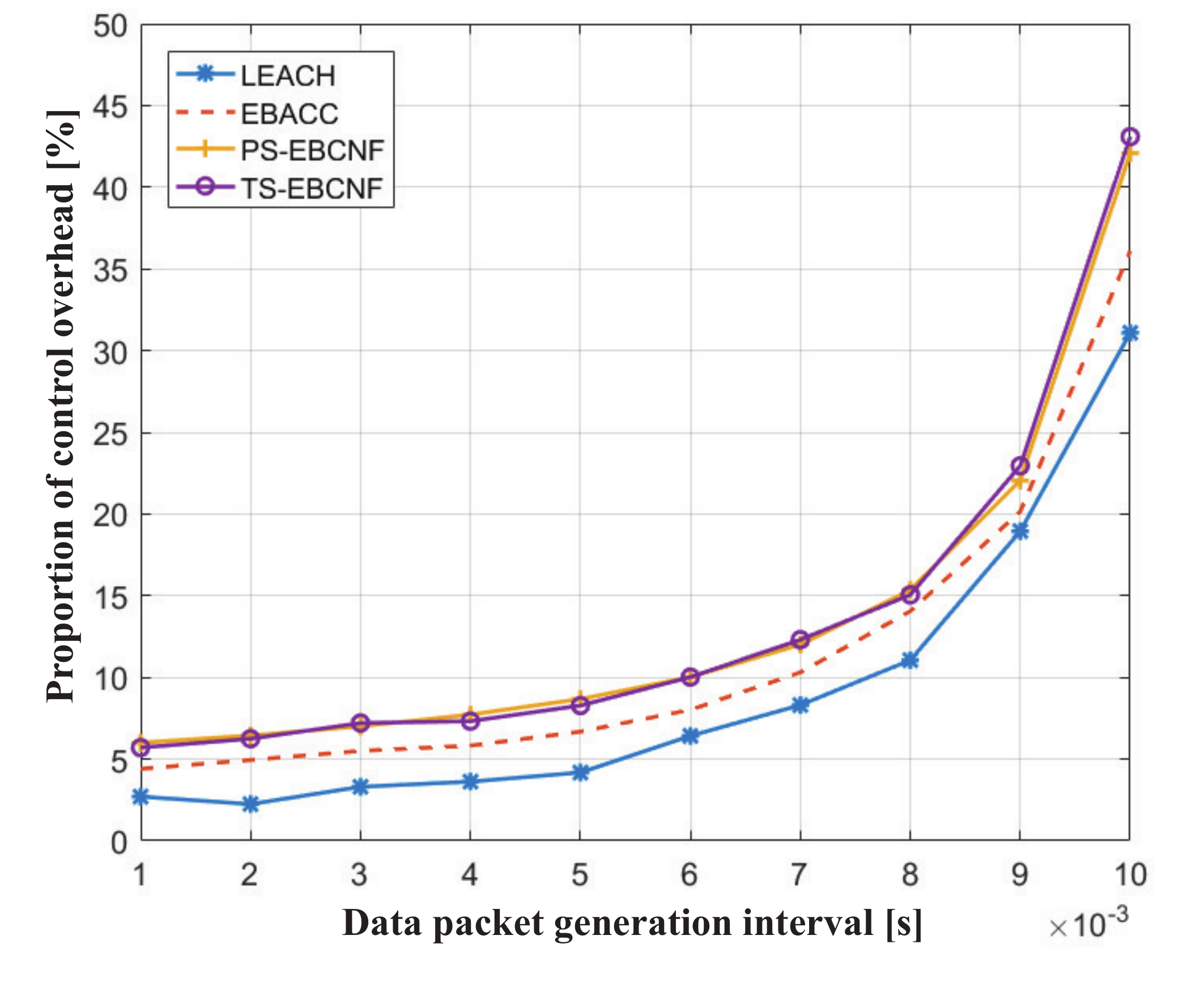}
	\caption{Comparison of control overhead.}
	\label{fig11}
\end{figure}
\FloatBarrier

	\end{itemize}

	\section{Conclusion}
For the limited energy of nano-nodes and the large-scale 
network and small communication range of WNSNs, we 
propose an energy-balanced clustering network framework 
EBCNF based on SWIPT. The framework uses an uneven 
clustering algorithm EBACC, which balances energy 
consumption within and between clusters. At the same time, we 
establish a SWIPT model to charge nano-nodes through a 
wireless energy carrying mechanism to break through the 
limited energy of nano-nodes. And the maximum-minimum 
algorithm is used to optimize the segmentation coefficient in 
the SWIPT mechanism to further improve the performance of 
SWIPT in the network. Finally, simulations results verify that 
the EBCNF framework has greater advantages over LEACH in 
terms of balancing network energy consumption, data 
transmission success rate and average throughput, and can be 
used as an effective routing framework for WNSNs.

	
	\bibliographystyle{elsarticle-num}
	\bibliography{refs}
	
	
	%
	%
	%

\end{document}